# Detecting Hollow Electron Beams via Smith-Purcell Radiation from a Metasurface

Daria Yu. Sergeeva[a], Dmitry V. Karlovets[b], and Alexey A. Tishchenko[a,c,*]

[a] *National Research Nuclear University "MEPhI", Kashirskoe sh. 31, Moscow, 115409, Russia*
[b] *School of Physics and Engineering, ITMO University, 197101 St. Petersburg, Russia*
[c] *Belgorod National Research University, Pobedy St. 85, Belgorod, 30801, Russia*

Hollow electron beams are today highly desirable for many applications, but are still challenging in view of their detection. In this Letter, we focus on the unique character of the electromagnetic radiation that relativistic hollow electron beams can produce when travelling above a metasurface. We investigate theoretically the specific features of the radiation in a coherent mode, which provides the highest intensity, and show that the radiation from a hollow beam can be considerably more intense than that from a conventional solid beam. This solves the problem of distinguishing between hollow and solid beams. Moreover, we consider the two-layer internal structure of a hollow beam and reveal that the radiation characteristics are sensitive to the width and population of each layer. This allows detecting the internal structure of hollow beams. Interestingly, we found that the factor describing the annular beam form is a separated multiplier in a conventional form factor, independent of the properties of periodic structure. Thus, we can conclude that our results will stay correct for different profiles of periodic structures and metasurfaces made of metaatoms of different topologies and forms. The results pave the way towards a variety of newly emerging applications based on hollow electron beams, very diverse in topics, such as manipulation of objects at the nano-level, studies of chiral matter, plasma acceleration in donut wakefields and even applications in huge facilities such as LHC for controlling proton beam halos etc.

*Introduction.* – Hollow electron beams are beams that are hollow inside, so they have a rectangular cross-sectional profile. They are significantly more stable [1, 2] than conventional solid electron beams in respect with the space charge effect due to the Coulomb repulsion between electrons.

Hollow beams are the basis for hollow electron lenses designed to control the halo of proton beams at the LHC and similar facilities [3–5]. Such a lens is a ring-shaped electron beam, and for it is crucially important to have an ideal annular transverse distribution with an electron distribution inside it close to rectangular [1, 6]. Such beams are generated, e.g., due to thermionic emission in a ring cathode [6].

Another application of hollow beams is the acceleration of positrons to high energies: this is possible in the wake fields formed when hollow electron beams move in plasma. This idea was confirmed by the results of computer simulation of the formation of donut wakefields and the dynamics of hollow electron beams [7–9]. Interestingly, even the creation of ultrashort attosecond electron beams (technologies just emerging today) can be based on hollow beams generated using nanofibers [10].

One more area in which hollow beams occur is twisted beams, i.e. beams of particles carrying orbital angular momentum (OAM). Twisted beams are promising for designing optical tweezers that manipulate objects at the nanoscale [11, 12], studying OAM-enhanced chirality [12, 13], inducing multiple transitions in atomic and nuclear physics [14], etc., and these beams are also essentially hollow.

The breadth of applications requires developed methods for diagnosing hollow beams. Recently, a new monitor based on fluorescence due to the interaction between electrons with supersonic gas, promising for LHC operation, was proposed to measure the two-dimensional profile of an electron beam [15, 16]. Yet, the results are obtained for deeply nonrelativistic electrons (7 keV), and the beams are very long, up to 25 $\mu s$ (more than a kilometer), which limits the applicability to only incoherent radiation mode. Short beams emit in a coherent mode, and methods based on fluorescence, as in Refs. [15, 16] is not valid. Also, it was shown [17] that transition radiation in the coherent mode is not suitable for distinguishing between solid and hollow beams, due to the similarity of radiation characteristics.

The question arises: how to diagnose the internal structure of the beam in practically important case when the beams are short and relativistic? Obviously, the answer must be based on the effect of the beam internal structure. What radically distinguishes hollow beams from conventional solid ones is their internal transverse structure. For this difference to appear, it is necessary to consider the radiation process sensitive to the transverse structure of the beam.

In Ref. [18] it was shown that the transverse distribution of the electrons is manifested in Smith-Purcell radiation (SPR), i.e. the spontaneous emission of electron beams travelling above the periodic structure. Thus, it is of interest to consider SPR from a hollow beam, and to see how the radiation characteristics depend on its internal structure. Going ahead, in this Letter we show that radiation intensities of a hollow and a conventional solid electron beam can, under realistic conditions, differ significantly.

*Smith-Purcell radiation from a metasurfaces.* – When a hollow electron beam passes near a periodic structure, the Smith-Purcell radiation is emitted as the result of dynamical polarization of matter by the Coulomb field of the moving electrons. Let us consider a metasurface being a flat monolayer consisting of $N_p$ periodically arranged particles. The particles are of sub-wavelength size, but arbitrary in the forms and topology, see the insertion in Fig. 1.

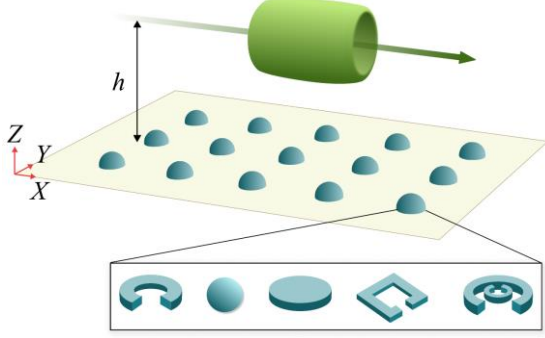

FIG. 1. A hollow beam travels above an array of metaatoms. In insertion, the different types of possible metaatoms are shown.

*The profile of hollow beams.* – The hollow electron beam can have uniform transverse distribution of the electrons, while the longitudinal one can be arbitrary. If the population of the beam is not extremely high and one can neglect the correlations between single electrons, the function of distribution of the electrons inside the beam reads

$$f(\mathbf{r}_{el}) = f_{tr}(y_{el}, z_{el}) f_l(x_{el}), \qquad (1)$$

with $\mathbf{r}_{el}$ being the radius-vector of electron relative to the center of the beam, $f_l$ describes the longitudinal distribution, $f_{tr}$ describes the transverse distribution; the beam moves along *x*-axis.

For the hollow beam with uniform distribution

$$f_{tr} = \begin{cases} 0, & \text{inside the beam} \\ \dfrac{1}{\pi(R_{out}^2 - R_{in}^2)}, & \text{outside} \end{cases} \qquad (2)$$

Here $R_{in}$ is the internal radius of the beam, $R_{out}$ is the outer one. The function $f(\mathbf{r}_{el})$ is normalized to unity.

Yet, the hollow beams can consist of two layers with a close to the uniform distribution inside [6], and in this case we have more complicated distribution function:

$$f_{tr} = \begin{cases} 0, & \text{outside the beam} \\ \dfrac{n}{\pi(R_m^2 - R_{in}^2)}, & \text{inside the inner ring} \\ \dfrac{1-n}{\pi(R_{out}^2 - R_m^2)}, & \text{inside the outer ring} \end{cases} \qquad (3)$$

where $R_m$ is the middle radius of the beam, see Fig. 2. Eq. (3) coincides with Eq. (2) when

$$\{R_{in} = R_m, n = 0\}, \{R_{out} = R_m, n = 1\}: \qquad (4)$$

we have a single-layer hollow beam.

The population parameter $n$ defines the part of the electrons being in the internal volume. E.g., this parameter is calculated to equal $n = 0.135$ for the parameters of the Ref. [6]: the total population of the beam, the current, radii of the both rings indicated in Ref. [6]. The calculated $n$ describes the profile of the beam with high precision, compare red (our) and dotted (taken from Ref. [6]) lines in Fig. 2.

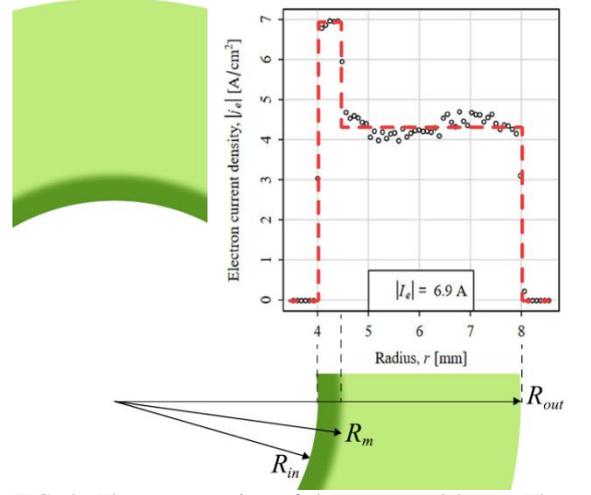

FIG. 2. The cross-section of the composed beam. The red line on the radius-dependence is described by Eq. (3) with parameter $n = 0.135$ selected so that to coincide with the dotted line taken from the Ref. [6].

*Intensity of radiation.* – The radiated energy per unit frequency and per unit solid angle, which for shortness we will call an intensity, is defined by the Fourier-transformed field of radiation at far distances:

$$I \equiv \left\langle \frac{dW(\mathbf{n},\omega)}{d\Omega d\hbar\omega} \right\rangle = \frac{cr^2}{\hbar} \left\langle |\mathbf{E}(\mathbf{r},\omega)|^2 \right\rangle, \qquad (5)$$

where $\mathbf{n}$ is the unit wave-vector of radiation, $\Omega$ is the solid angle of observation of radiation, $\omega$ is the radiation frequency, the angle brackets stand for averaging over the positions of all the electrons in the beam, $c$ is the speed of light in vacuum, $\hbar$ is the Planck constant, $r$ is the distance to the point of observation, $\mathbf{E}(\mathbf{r},\omega)$ is the field of radiation.

At far distances $r$ ($kr \gg 1$, where $k = \omega/c$) the field of radiation is expressed through the Fourier-image of current density $\mathbf{j}(\mathbf{k},\omega)$:

$$\mathbf{E}(\mathbf{r},\omega) = -i\frac{(2\pi)^3}{\omega} \frac{e^{ikr}}{r} \left(\mathbf{k} \times (\mathbf{k} \times \mathbf{j}(\mathbf{k},\omega))\right). \qquad (6)$$

The polarization currents are induced at the metaatoms a metasurface consists of. As the metaatoms are sub-wavelength in their size, the current density is described by the sum of dipole moments

$$\mathbf{j}(\mathbf{r},t) = -i\omega \sum_p \mathbf{d}(\mathbf{r}_p,\omega) \delta(\mathbf{r} - \mathbf{r}_p),$$

$$\mathbf{d}(\mathbf{r}_p,\omega) = \alpha(\omega) \sum_{el=1}^N \mathbf{E}_{el}(\mathbf{r}_p,\omega), \qquad (7)$$

$N$ is the number of electrons in the beam, $\alpha(\omega)$ is the polarizability of the particle, which characterizes how the particle reacts to the external field, $\mathbf{E}_{el}(\mathbf{r}_p,\omega)$ is the electron's Fourier-transformed Coulomb field at the point with the radius-vector $\mathbf{r}_p$:

$$\mathbf{E}_{el}(\mathbf{r}_p, \omega) = -\frac{ie\omega}{\pi v^2 \gamma} e^{i\frac{\omega}{v}(x_p - x_{el})}$$
$$\times \left\{ \frac{\mathbf{v}}{v\gamma} K_0\left(\frac{\omega\rho}{v\gamma}\right) + i\frac{\boldsymbol{\rho}}{\rho} K_1\left(\frac{\omega\rho}{v\gamma}\right) \right\}, \quad (8)$$

where $e$ is the electron's charge, $\gamma = 1/\sqrt{1-\beta^2}$ is the Lorentz-factor of the electrons, $v$ is the velocity of the electrons, $K_0$ and $K_1$ are modified Bessel functions of zero-th and first orders, and the following vector is introduced

$$\boldsymbol{\rho} = (y_p - y_{el})\mathbf{e}_y - (z_{el} + h)\mathbf{e}_z, \quad (9)$$

with $\mathbf{e}_y$ and $\mathbf{e}_z$ being the base vectors, $h$ being the impact-parameter of the beam, i.e. the shortest distance between its center and the plane of the grating.

Eq. (5) combined with Eqs. (6)-(9) gives

$$I = \frac{\omega^2}{c\hbar} |\alpha(\omega)|^2 \left\langle \left| \sum_{el=1}^{N} \sum_{p} (\mathbf{k} \times \mathbf{E}_{el}(\mathbf{r}_p, \omega)) e^{-i\mathbf{k}\mathbf{r}_p} \right|^2 \right\rangle. \quad (10)$$

The theory constructed here is based on consideration of the particles as non-interacting, following the approach developed in [19–21]. The collective effects caused by the interaction of single metaatoms can manifest themselves near resonance conditions, see, e.g., [22–26]. Outside of resonance effects, a theory that neglects collective effects is valid in practice, which is confirmed by good agreement with experiment [20].

The averaging in Eq. (10) reduces to integrating the function inside the angle brackets over the position of electrons inside the beam. Considering coherent radiation, which is approximately $N$ times more intense than incoherent, for $N \gg 1$ we can keep only off-diagonal terms in the squared sum in Eq. (10).

Now, using Eq. (8) and following the approach from Ref. [18], we obtain from Eqs. (10) and (3)

$$I_{coh} = N^2 \frac{\omega^2}{\hbar c} |\alpha(\omega)|^2 F_l \left| \sum_{p} e^{-i\mathbf{k}\mathbf{r}_p} e^{ix_p \omega/v} \mathbf{k} \times \mathbf{A} \right|^2. \quad (11)$$

Here

$$F_l = \left| \int dx_{el} f_l(x_{el}) e^{-ix_{el}\omega/v} \right|^2, \quad (12)$$

is the longitudinal form factor of the beam and

$$\mathbf{A} = \mathbf{A}_{in} + \mathbf{A}_{out}, \quad (13)$$

$$\mathbf{A}_{in} = -\frac{ie\omega}{\pi v^2 \gamma} \iint_{S_{in}} dy_{el} dz_{el} \frac{n}{\pi(R_m^2 - R_{in}^2)}$$
$$\times \left\{ \frac{\mathbf{v}}{v\gamma} K_0\left(\frac{\omega\rho}{v\gamma}\right) + i\frac{\boldsymbol{\rho}}{\rho} K_1\left(\frac{\omega\rho}{v\gamma}\right) \right\}, \quad (14)$$

$$\mathbf{A}_{out} = -\frac{ie\omega}{\pi v^2 \gamma} \iint_{S_{out}} dy_{el} dz_{el} \frac{1-n}{\pi(R_{out}^2 - R_m^2)}$$
$$\times \left\{ \frac{\mathbf{v}}{v\gamma} K_0\left(\frac{\omega\rho}{v\gamma}\right) + i\frac{\boldsymbol{\rho}}{\rho} K_1\left(\frac{\omega\rho}{v\gamma}\right) \right\}, \quad (15)$$

with $S_{in}$, $S_{out}$ being the internal and outer cross-sectional areas of the beam.

Depending on an explicit form of the function $f_l$, the longitudinal form factor $F_l$ can be anyone. For a constant uniform, Gaussian and modulated distributions this factor was calculated in [18, 27–29]. For short beams with $l \ll \lambda$, $F_l = 1$ with high accuracy.

Integration in Eq. (14) is performed in the following way. First, let us represent the integral $\mathbf{A}_{in}$ in the form

$$\mathbf{A}_{in} = -\frac{ie}{2\pi^2 v} \iint_{S_{in}} \frac{dy_{el} dz_{el}}{\pi(R_m^2 - R_{in}^2)} n \iint dq_y dq_z$$
$$\times e^{iq_y(y_p - y_{el})} e^{-iq_z(h+z_{el})} \frac{\frac{\omega}{v\gamma^2}\mathbf{e}_x + q_y\mathbf{e}_y + q_z\mathbf{e}_z}{q_y^2 + q_z^2 + \frac{\omega^2}{v^2\gamma^2}}. \quad (16)$$

Due to cylindrical symmetry of the problem, it is convenient to calculate the integrals in polar coordinates:

$$\iint_{S_{in}} dy_{el} dz_{el} e^{-iq_y y_{el}} e^{-iq_z z_{el}} =$$
$$= \int_0^{2\pi} d\varphi \int_{R_{in}}^{R_m} dR e^{-iq_y R\cos\varphi} e^{-iq_z R\sin\varphi} R = \quad (17)$$
$$= 2\pi \frac{R_m J_1(QR_m) - R_{in} J_1(QR_{in})}{Q},$$

where $J_1$ is the Bessel function of the first order and the polar system for $q_{y,z}$ was introduced:

$$q_y = Q\cos\phi, \quad q_z = Q\sin\phi. \quad (18)$$

Eq. (16) can be represented in the form

$$\mathbf{A}_{in} = -\frac{ie}{\pi^2 v} n \frac{R_m \mathbf{B}(R_m) - R_{in} \mathbf{B}(R_{in})}{R_m^2 - R_{in}^2}, \quad (19)$$

$$\mathbf{B}(R) = \left[\frac{\omega}{v\gamma^2}\mathbf{e}_x - i\mathbf{e}_y \frac{\partial}{\partial y_p} + i\mathbf{e}_z \frac{\partial}{\partial h}\right]$$
$$\times \int_0^{2\pi} d\phi \int_0^{+\infty} dQ e^{iQy_p \cos\phi} e^{-iQh\sin\phi} \frac{J_1(QR)}{Q^2 + (\omega/v\gamma)^2}. \quad (20)$$

Integrating over $\phi$, we obtain

$$\mathbf{B}(R) = 2\pi \left[\frac{\omega}{v\gamma^2}\mathbf{e}_x - i\mathbf{e}_y \frac{\partial}{\partial y_p} + i\mathbf{e}_z \frac{\partial}{\partial h}\right] \times$$
$$\times \int_0^{+\infty} dQ \frac{J_1(QR) J_0(Q\rho_0)}{Q^2 + (\omega/v\gamma)^2}, \quad (21)$$

where $\rho_0 = \sqrt{y_p^2 + h^2}$, $h$ is an impact-parameter, i.e. the shortest distance between the beam trajectory and the metasurfaces, see Fig. 1. Using [30], we find

$$\int_0^{+\infty} dQ \frac{J_1(QR) J_0(Q\rho_0)}{Q^2 + (\omega/v\gamma)^2} =$$
$$= \frac{v\gamma}{\omega} I_1\left(\frac{R\omega}{v\gamma}\right) K_0\left(\frac{\rho_0 \omega}{v\gamma}\right). \quad (22)$$

After calculation of the derivatives we obtain:

$$\mathbf{B}(R) = 2\pi I_1\left(\frac{R\omega}{v\gamma}\right) \times$$
$$\times \left[\frac{\mathbf{e}_x}{\gamma}K_0\left(\frac{\rho_0\omega}{v\gamma}\right) + i\frac{\mathbf{\rho}_0}{\rho_0}K_1\left(\frac{\rho_0\omega}{v\gamma}\right)\right], \quad (23)$$

$\mathbf{\rho}_0 = y_p\mathbf{e}_y - h\mathbf{e}_z$, and, finally,

$$\mathbf{A}_{in} = -\frac{ie}{\pi v}\left[\frac{\mathbf{e}_x}{\gamma}K_0\left(\frac{\rho_0\omega}{v\gamma}\right) + i\frac{\mathbf{\rho}_0}{\rho_0}K_1\left(\frac{\rho_0\omega}{v\gamma}\right)\right]$$
$$\times 2n\frac{R_m I_1\left(\frac{R_m\omega}{v\gamma}\right) - R_{in} I_1\left(\frac{R_{in}\omega}{v\gamma}\right)}{R_m^2 - R_{in}^2}. \quad (24)$$

Similarly,

$$\mathbf{A}_{out} = -\frac{ie}{\pi v}\left[\frac{\mathbf{e}_x}{\gamma}K_0\left(\frac{\rho_0\omega}{v\gamma}\right) + i\frac{\mathbf{\rho}_0}{\rho_0}K_1\left(\frac{\rho_0\omega}{v\gamma}\right)\right]$$
$$\times 2(1-n)\frac{R_{out} I_1\left(\frac{R_{out}\omega}{v\gamma}\right) - R_m I_1\left(\frac{R_m\omega}{v\gamma}\right)}{R_{out}^2 - R_m^2}. \quad (25)$$

Thus, Eq. (11) with Eqs. (24), (25) describes the radiation of any of the beams considered here: single-layer hollow, two-layer, and a solid one. Now we can answer the question, what is the difference between the radiation of a hollow and a solid beam?

*Radiation of a single-layer hollow beam* – Thus, in the special case of a single-layer when Eq. (4) is satisfied, Eq. (11) reads

$$I_{coh} = N^2 I_{single} F_{tr} F_l, \quad (26)$$

where $I_{single}$ is the intensity of radiation from a single electron, which is placed at the very center of the beam:

$$I_{single} = \frac{1}{137}\frac{\omega^4}{\pi^2 v^4 \gamma^2}|\alpha(\omega)|^2$$
$$\times \left|\mathbf{k}\times\sum_p\left(\frac{\mathbf{v}}{\gamma v}K_0\left(\frac{\omega\rho_0}{v\gamma}\right) + i\frac{\mathbf{\rho}_0}{\rho_0}K_1\left(\frac{\omega\rho_0}{v\gamma}\right)\right)e^{-i\mathbf{k}\mathbf{r}_p}e^{ix_p\omega/v}\right|^2,$$
(27)

and

$$F_{tr} = \frac{4I_1^2(x)}{x^2}G, \quad (28)$$

is the transverse form factor of the beam, with $I_1$ being the modified Bessel function of the first order and $x = \omega R_{out}/v\gamma$. The factor

$$G = \frac{1}{(1-\mu^2)^2}\left(1 - \frac{\mu I_1(\mu x)}{I_1(x)}\right)^2, \quad (29)$$

characterizes how the beam is populated by the electrons; $\mu = R_{in}/R_{out}$. It is the factor $G$ that describes the difference between the radiations from hollow and solid beams.

For $R_{in} = 0$ the factor $G$ in Eq. (29) goes to unity, and transverse form factor is

$$F_{tr} \to 4I_1^2(x)/x^2, \quad (30)$$

which exactly coincides with the form factor obtained in [28] for the cylindrical beam with the uniform distribution with the radius $R_{out}$. Thus, we see that the form factor of the hollow beam differs from that for conventional, i.e. not-hollow beam, by the factor $G$. As is seen in Fig. 3, this is a smooth function, which achieves minimum equal to 1 at $\mu x = 0$, and achieves maximum equal to $(1-\mu)^{-2}$ at $x \to \infty$.

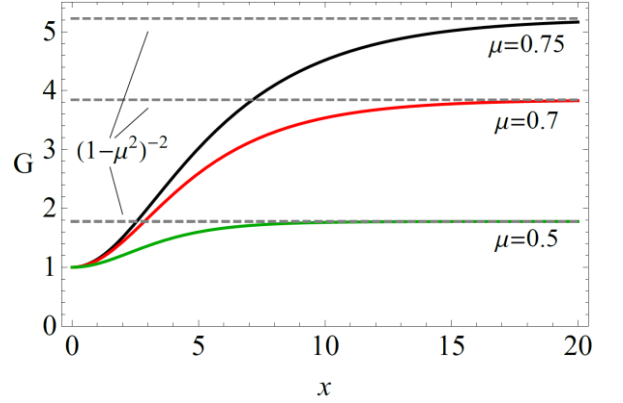

FIG. 3. Factor $G$ that defines the difference between the radiations from hollow and solid beams, see Eq. (29), depending on $x$ for different values of $\mu$.

Because of the evident condition $R_{out} > R_{in}$, we always have $\mu < 1$.

Why a hollow beam can shine brighter than a solid one? We assume that the origin of this effect is clear: compared with a solid beam, in a hollow one part of electrons is closer to the target, i.e. the Coulomb field of the beam exciting the radiation is stronger.

The variable $x$ comprises the frequency $\omega$, so Fig. 3 shows the spectral properties as well. Yet, the factor $G$ is only a part of the full intensity given by Eq. (26), the other parts of which are also dependent on the frequency and other parameters. Fig. 4 shows the spectrum, $\nu = \omega/2\pi$.

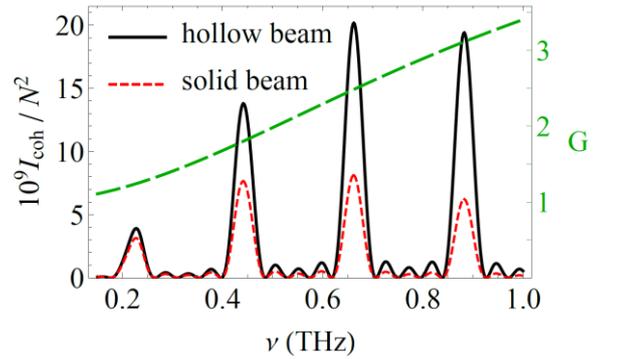

FIG. 4. Radiation intensity for hollow (black) and solid (red) beams with equal outer radius. The green line corresponds to the factor $G$. Here $\gamma = 16$, the angle of observation $\theta = 30$ degree $(k_x = k\cos\theta)$, the grating $5\times 5$ consists of spherical particles of the radius 0.1 mm with dielectric permittivity of the material $\varepsilon = 2.1$, the periods of the grating are $d_x = d_y = 10$ mm, $h = 6$ mm, $R_{out} = 4.5$ mm, $\mu = 0.75$.

It is seen that for the hollow beam each diffraction order is enhanced, but with different factor: the right peak is three times higher than the red one, while the left peak is almost unchanged; this law is determined by the form of the curve in Fig. 3. This allows one to measure the internal radius of the beam $R_{in}$. Namely, having measured values of the maxima of at least three peaks shown in Fig. 4 (black lines), it is necessary to divide them by the values of the single-particle intensity determined by Eq. (27) multiplied by $4I_1^2(x)/x^2$, and then draw a curve of the same shape as in Fig. 3 along three points, thereby determining the value of $\mu$, subject to fixing other parameters of the beam.

Now let us consider two different cases.

If $R_{out} \ll \gamma\beta\lambda/2\pi$, then $x \ll 1$ and $\mu x \ll 1$, and, consequently, $G \simeq 1$. Here Eq. (28) reads

$$F_{tr} = \frac{4I_1^2(x)}{x^2}G \simeq 1. \qquad (31)$$

This means that under conditions of high spatial coherence of radiation there is no difference between the radiation of a hollow beam and a solid one.

If $R_{in} \gg \gamma\beta\lambda/2\pi$, i.e. $\mu x \gg 1$, then $x \gg 1$ and

$$G \simeq (1-\mu^2)^{-2}. \qquad (32)$$

Here $\mu = R_{in}/R_{out} < 1$, but with increasing $\mu$ the factor $G$ also increases: e.g., $G = 100$ for $\mu = 0.9$; for less $\mu$ see Fig. 3.

Does it mean that in the case of $\mu x \gg 1$ we can increase the radiation from a hollow beam unlimitedly, operating its internal structure? No. Actually, since $h > R_{out} > R_{in}$, then for $R_{in} \gg \gamma\beta\lambda/2\pi$ we have

$$h > \gamma\beta\lambda/2\pi. \qquad (33)$$

Under this condition the interaction between the Coulomb field of the electrons and the target is hardly effective, which mathematically is determined by the decreasing behavior of the functions $K_0$ and $K_1$ for large arguments. Therefore, for $\mu x \gg 1$ we can really observe the clear difference (see Fig. 3) in the intensity of radiation of hollow and usual beams, but only in regions of parameters described by Eq. (33). The maximum difference from solid beams is shown by the hollow beams in which electrons are maximally shifted to the outer radius of the beam.

*Radiation of a two-layer hollow beam* – In more general case of a two-layer hollow beam, see Fig. 2, Eqs. (26), (27) stay correct, while Eq. (28) reads

$$F_{tr} = \frac{4I_1^2(x)}{x^2}G', \qquad (34)$$

with

$$G' = \left| n\frac{\mu_m I_1(x\mu_m) - \mu I_1(x\mu)}{(\mu_m^2 - \mu^2)I_1(x)} + (1-n)\frac{I_1(x) - \mu_m I_1(x\mu_m)}{(1-\mu_m^2)I_1(x)} \right|^2, \qquad (35)$$

where $\mu_m = R_m/R_{out}$. For $x\mu \to \infty$ and $x\mu_m \to \infty$ we get instead of Eq. (35)

$$G' = \frac{(1-n)^2}{(1-\mu_m^2)^2}. \qquad (36)$$

For $R_m = R_{out}$ we get $\mu_m = 1$ and $n = 1$, and Eq. (35) transforms into Eq. (29), while Eq. (36) comes into $G' = G = (1-\mu^2)^{-2}$, as expected, due to $\mu_m \to \mu$.

Fig. 5 demonstrates the factor $G'$, defining the difference between radiation of a two-layer hollow and a solid beams.

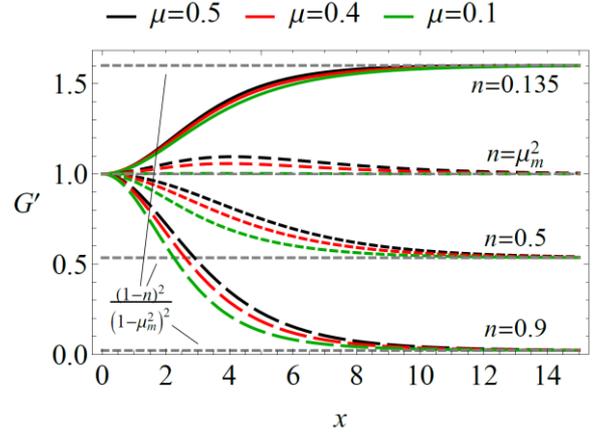

Fig. 5. Here, as in Ref. [6], $\mu_m = 9/16$ and the green curve is plotted for $n = 0.135$.

*Discussion and outlook.* – As shown above, the difference between the radiation of hollow and solid beams does not depend on the properties of periodic structure. Actually, the explicit form of the intensity given by Eqs. (26) - (29), including Eqs. (34), (35) for two-layer hollow beams, factorizes into individual factors defining the radiation from a single electron, see Eq. (27), and the production of two form factors: a longitudinal (arbitrary) and a transverse one (calculated above). And this is the transverse form factor that contains the factor $G$ (or $G'$) determining the whole dependence of the radiation intensity on the variable $\mu = R_{in}/R_{out}$ (and $\mu_m = R_m/R_{out}$ in the case of two layers) which, in its turn, defines the crossection of a hollow beam. Thus, we can conclude that the method diagnosing the internal structure of the beams, allowing one to distinguish hollow beams from solid ones, is correct for any kind of grating.

We can already use our results to draw general conclusions about the characteristics of radiation from hollow beams. Another useful output of this Letter will be to extend the solution to detect the internal structure of hollow beams, including the twisted ones (OAM-beams), thus providing an immediate impact in the numerous prospective fields of applications of the twisted charged particles and their beams.

Recently, the important role of twisted electrons in revealing the quantum wave nature of free electrons through their spontaneous emission using Smith-Purcell effect has been actively discussed [31–36]. These papers focus only on single electrons, but the electrons are described in terms of wave packets, which are similar to electron beams due to the statistical interpretation of the wave function.

Moreover, when the electrons are relativistic, electron beams rather than single electrons are used in practice. Therefore, although above we did not use the formalism of Bessel or Gauss-Laguerre beams, often used for twisted electrons, we can cautiously assume that our results pave the way to their diagnostics.

The study was supported by the Foundation for the Advancement of Theoretical Physics and Mathematics "BASIS": D.S. is grateful for the support under grant 23-1-3-2-1, and D.K. – under grant 22-1-2-64-1.

________________

[*]tishchenko@mephi.ru